\documentstyle[12pt]{article}
\newcommand{\Cslash}{\not \!\! C}
\newcommand{\Dslash}{\not \!\! D}
\begin{document}
\large
\begin{flushright}{ UT-844\\
April ,1999\\
}
\end{flushright}
\vfil

\begin{center}
{\large{\bf Relation $Tr \gamma_{5}= 0$ and the index theorem in lattice gauge 
theory }}
\end{center}
\vskip .5 truecm
\centerline{\bf Kazuo Fujikawa}
\vskip .4 truecm
\centerline {\it Department of Physics,University of Tokyo}
\centerline {\it Bunkyo-ku,Tokyo 113,Japan}
\vskip 0.5 truecm

\makeatletter
\@addtoreset{equation}{section}
\def\theequation{\thesection.\arabic{equation}}
\makeatother

\normalsize
\begin{abstract}
 The relation $Tr \gamma_{5}= 0$ implies the contribution to the trace from unphysical (would-be) species doublers in lattice gauge theory. This statement is also true for the Pauli-Villars regularization in continuum theory. If one insists on $Tr \gamma_{5}= 0$, one thus inevitably includes unphysical states in the  Hilbert space.  If one truncates the trace to the contribution from physical species only, one obtains  $\tilde{T}r \gamma_{5} = n_{+} - n_{-}$ which is equal to the Pontryagin index. A smooth continuum limit of  $\tilde{T}r \gamma_{5} = Tr \gamma_{5}(1-(a/2)D) = n_{+} - n_{-}$ for the Dirac operator $D$ satisfying the Ginsparg-Wilson relation leads to the natural treatment of chiral anomaly in continuum path integral. In contrast, the continuum limit of  $Tr \gamma_{5}= 0$ is not defined consistently. It is shown that the non-decoupling of heavy fermions in the anomaly 
calculation is crucial to understand the consistency of the customary lattice calculation of anomaly where  $Tr \gamma_{5}= 0$ is used. We also comment on  a closely related phenomenon in the analysis of the photon phase operator where the notion of  index and the modification of index by a finite cut-off play a crucial role.
 
\end{abstract}

\newpage
\section{Introduction}
Recent developments in the treatment of  fermions in lattice gauge theory led  to a better understanding of chiral symmetry not only in lattice theory [1]-[7]but possibly also in continuum theory[8]. These developments are based on a  
hermitian lattice Dirac  operator $\gamma_{5}D$ which satisfies  the so-called Ginsparg-Wilson relation[1]
\begin{equation}
\gamma_{5}D + D\gamma_{5} = aD\gamma_{5}D.
\end{equation}
 An explicit example of the operator satisfying (1.1) and free of species doubling has been given by Neuberger[2]. The operator has also been discussed as a fixed point  form of block transformations [3].
The relation (1.1) led to the  interesting analyses of the notion of  index in lattice gauge theory[4]-[9]. Here  $\gamma_{5}$ is a hermitian chiral Dirac matrix. 

The index relation is generally written as [4][5]
\begin{equation}
 Tr \gamma_{5}(1 -\frac{1}{2}aD) = n_{+} - n_{-}
\end{equation}
which is confirmed by [8] 
\begin{eqnarray}
Tr [ \gamma_{5}(1-\frac{1}{2}aD)]&=& \sum_{n}\{\phi^{\dagger}_{n}\gamma_{5}\phi_{n} - \frac{1}{2}\phi^{\dagger}_{n}\gamma_{5}aD\phi_{n}\}\nonumber\\
&=& \sum_{ \lambda_{n}=0}\phi^{\dagger}_{n}\gamma_{5}\phi_{n} + \sum_{ \lambda_{n}\neq 0}\phi^{\dagger}_{n}\gamma_{5}\phi_{n} - \sum_{n}\frac{1}{2}a\lambda_{n}\phi^{\dagger}_{n}\phi_{n}\nonumber\\
&=&\sum_{ \lambda_{n}=0}\phi^{\dagger}_{n}\gamma_{5}\phi_{n}\nonumber\\
&=& n_{+} - n_{-} =  index
\end{eqnarray}
where $n_{\pm}$ stand for the number of  normalizable zero modes in 
\begin{equation}
\gamma_{5}D\phi_{n}=\lambda_{n}\phi_{n}
\end{equation}
for the {\em hermitian}  operator $\gamma_{5}D$ with simultaneous eigenvalues $\gamma_{5}\phi_{n}= \pm \phi_{n}$.  
We also used the relation  
\begin{equation}
\phi^{\dagger}_{n}\gamma_{5}\phi_{n} = \frac{a}{2}\lambda_{n}\phi^{\dagger}_{n}\phi_{n} = \frac{a}{2}\lambda_{n}
\end{equation}
for $\lambda_{n}\neq 0$, which is derived by sandwiching the relation(1.1) by 
$\phi^{\dagger}_{n}\gamma_{5}$ and $\phi_{n}$. It should be emphasized that 
the relation (1.3) is derived {\em without} using $Tr \gamma_{5}=0$. The inner product $\phi^{\dagger}_{n}\phi_{n} = (\phi_{n},\phi_{n})
\equiv \sum_{x} a^{4}\phi^{\star}_{n}(x)\phi_{n}(x)$ is defined by summing over all the lattice points, which are not explicitly written
in $\phi_{n}$. See Appendix for further notational details.

An advantage of  gauge theory defined on a finite lattice is that one can analyze some subtle aspects of chiral symmetry in continuum theory in a well-defined finite setting. The purpose of the present note is to study some of those aspects of chiral symmetry in the hope that this analysis also deepens our understanding of lattice regularization. In the  path integral treatment of chiral anomaly in continuum, the relation
\begin{equation}
Tr \gamma_{5}  = n_{+} - n_{-}
\end{equation}
in a suitably regularized sense plays a fundamental role[10][11]. On the other hand,
it is expected that the relation
\begin{equation}
Tr \gamma_{5} = 0
\end{equation}
holds on a finite lattice. As  Chiu pointed out[12], this relation (1.7) leads to 
an interesting {\em constraint} 
\begin{equation}
Tr \gamma_{5} = n_{+}- n_{-} + N_{+} - N_{-} = 0  
\end{equation} 
where $N_{\pm}$ stand for the number of eigenstates $\gamma_{5}D\phi_{n}=\pm (2/a)\phi_{n}$  with  $\gamma_{5}\phi_{n}= \pm \phi_{n}$, respectively.
It is important to recognize that $Tr \gamma_{5} = 0$ means that this relation holds for {\em any} sensible basis set with any background gauge field in a given theory, which may be used to define
the trace. Consequently, the seemingly trivial relation $Tr \gamma_{5} = 0$
in fact carries important physical information. In this note we show that 
$  Tr \gamma_{5} = 0$ implies the inevitable 
contribution from unphysical (would-be) species doublers in lattice theory or 
an unphysical bosonic spinor in Pauli-Villars regularization. In other words, $Tr \gamma_{5} = 0$ cannot hold in the physical Hilbert space consisting of physical states only, and the continuum limit of $  Tr \gamma_{5} = 0$ is not defined consistently, as is seen in (1.8).  It is shown that the failure of the decoupling of heavy fermions in the anomaly calculation is crucial to
understand the consistency of  the customary lattice calculation of anomaly where $  Tr \gamma_{5} = 0$ is used.
( The continuum limit in this paper stands for the so-called ``naive''
continuum limit with $a\rightarrow 0$, and the lattice size is gradually extended to
infinity for any finite $a$ in the process of taking the limit $a\rightarrow 0$.)
  We then  discuss the possible implications of our analysis 
on the treatment of chiral anomalies in continuum theory. We also briefly
comment on an analogous phenomenon in the analysis of  photon phase operator, where the notion of index plays a crucial role. 

\section{Consistency of the relation $Tr \gamma_{5} = 0$ }
In the previous section we have seen  that the consistency of the relation $Tr \gamma_{5} = 0$ requires the presence of the $N_{\pm}$ states for an operator $\gamma_{5}D$ satisfying (1.1) on a finite lattice. We thus want to analyze the nature of the $N_{\pm}$ states in more detail. For this purpose, we start with the conventional Wilson operator $D_{W}$
\begin{eqnarray}
D_{W}(n,m)&\equiv&i\gamma^{\mu}C_{\mu}(n,m) + B(n,m) -\frac{1}{a}m_{0}\delta_{n,m},\nonumber\\
C_{\mu}(n,m)&=&\frac{1}{2a}[\delta_{m+\mu,n}U_{\mu}(m) - \delta_{m,n+\mu}U^{\dagger}_{\mu}(n)],\nonumber\\
B(n,m)&=&\frac{r}{2a}\sum_{\mu}[2\delta_{n,m}-\delta_{m+\mu,n}U_{\mu}(m)
-\delta_{m,n+\mu}U^{\dagger}_{\mu}(n)],\nonumber\\
U_{\mu}(m)&=& \exp [iagA_{\mu}(m)],
\end{eqnarray}
where we added a constant mass term to $D_{W}$ for later convenience. Our 
matrix convention is that $\gamma^{\mu}$ are anti-hermitian, $(\gamma^{\mu})^{\dagger} = - \gamma^{\mu}$, and thus $\not \!\! C\equiv \gamma^{\mu}C_{\mu}(n,m)$ is hermitian
\begin{equation}
\not \!\! C ^{\dagger} = \not \!\! C.
\end{equation}
Since the operator $\Cslash$ forms the basis for any fermion operator on the 
lattice, we start with the analysis of $\Cslash$.

\subsection{Operator $\Cslash$ and  $Tr \gamma_{5} = 0$}
It was noted elsewhere[8] that $Tr \gamma_{5} = 0$ implies the species doubling 
for the operator $\Cslash$. The basic reasoning is based on the index relation
\begin{equation}
dim\ ker\ (\frac{1- \gamma_{5}}{2})\Cslash(\frac{1+\gamma_{5}}{2}) - dim\ ker\ (\frac{1+\gamma_{5}}{2})\Cslash(\frac{1- \gamma_{5}}{2})=0
\end{equation}
where we understand $(\frac{1- \gamma_{5}}{2})\Cslash(\frac{1+\gamma_{5}}{2})$
as standing for the two-component operator $b$ in 
\begin{eqnarray}
\Cslash&=&\left(\begin{array}{cc}
          0&b^{\dagger}\\
          b&0 
          \end{array}\right).
\end{eqnarray}
This form of $\Cslash$ is deduced by noting $\Cslash^{\dagger} = \Cslash$ and
$\gamma_{5}\Cslash + \Cslash\gamma_{5}=0$ in the representation where $\gamma_{5} $ is diagonal. The operator $b(m,n)$ projects a two-component spinor on a
finite lattice to another two-component spinor on the same lattice , and thus it is a square matrix in the coordinate representation.
 For a general finite dimensional square matrix $M$, the index theorem $dim\ ker\ M - dim\ ker\ M^{\dagger} = 0$ holds[8], where $dim\ ker\ M $, for example,
stands for the number of normalizable modes in $Mu_{n}=0$. In the present context, $dim\ ker\ (\frac{1- \gamma_{5}}{2})\Cslash(\frac{1+\gamma_{5}}{2}) =
dim\ ker\ b $ stands for  the number of normalizable zero modes in 
\begin{equation}
\Cslash\phi_{n}= 0
\end{equation}
with $(\frac{1+\gamma_{5}}{2})\phi_{n}= \phi_{n}$. Thus the index relation (2.3) shows that possible zero modes with $\gamma_{5}\phi_{n}= \pm \phi_{n}$ are always paired. The eigenstates with non-zero eigenvalues 
in 
\begin{equation}
\Cslash\phi_{n}= \lambda_{n}\phi_{n}
\end{equation}
give a vanishing contribution to the trace $ Tr \gamma_{5}$ since 
\begin{equation}
\phi_{n}^{\dagger}\gamma_{5}\phi_{n} = 0
\end{equation}
by noting $\gamma_{5}\Cslash+ \Cslash\gamma_{5}=0$. The index relation (2.3) is thus equivalent to $ Tr \gamma_{5}=0$.
 
If one recalls the Atiyah-Singer index theorem[10][13] written in the same notation as (2.3)
\begin{equation}
dim\ ker\ (\frac{1- \gamma_{5}}{2})\Dslash(\frac{1+\gamma_{5}}{2}) - dim\ ker\ (\frac{1+\gamma_{5}}{2})\Dslash(\frac{1- \gamma_{5}}{2})=\nu  
\end{equation}
where $\nu$ stands for the Pontryagin index (i.e., an integral of anomaly) and 
$\Dslash \equiv \gamma^{\mu} (\partial_{\mu} - igA_{\mu})$, one sees that a smooth continuum
limit of the lattice index relation (2.3) for a general background gauge field configuration is {\em inconsistent} with the absence of species doublers.

In the present $\Cslash$, a very explicit construction of species doublers 
is known.  For a square lattice one can explicitly show that the simplest lattice fermion action
\begin{equation}
S = \bar{\psi}i\Cslash\psi
\end{equation}
is invariant under the transformation[14]
\begin{equation}
\psi^{\prime}= {\cal T}\psi,\ \bar{\psi}^{\prime}= \bar{\psi}{\cal T}^{-1}
\end{equation}
where ${\cal  T}$ stands for any one of the following 16 operators
\begin{equation}
1,\ T_{1}T_{2},\ T_{1}T_{3},\ T_{1}T_{4},\ T_{2}T_{3},\ T_{2}T_{4},\ T_{3}T_{4},\ T_{1}T_{2}T_{3}T_{4},
\end{equation}
and 
\begin{equation}
T_{1},\ T_{2},\ T_{3},\ T_{4},\ T_{1}T_{2}T_{3},\ T_{2}T_{3}T_{4},\ T_{3}T_{4}T_{1},\ T_{4}T_{1}T_{2}.
\end{equation}
The operators  $T_{\mu}$  are  defined by 
\begin{equation}
T_{\mu}\equiv \gamma_{\mu}\gamma_{5}\exp {(i\pi x^{\mu}/a)}  
\end{equation}
and  satisfy the relation
\begin{equation}
T_{\mu}T_{\nu} + T_{\nu}T_{\mu}=2\delta_{\mu\nu}
\end{equation}
with  $T_{\mu}^{\dagger} = T_{\mu} = T^{-1}_{\mu}$ for anti-hermitian $\gamma_{\mu}$. 
We denote the 16 operators by ${\cal T}_{n}, \ \ n=0\sim 15$, in the following 
with ${\cal T}_{0}=1$.
By recalling that the operator $T_{\mu}$ adds the  momentum $\pi/a$ to the 
fermion momentum $k_{\mu}$, we cover the entire Brillouin zone  
\begin{equation}
- \frac{\pi}{2a} \leq k_{\mu} <  \frac{3\pi}{2a}
\end{equation}
by the operation (2.10) starting with the free fermion defined in
\begin{equation}
- \frac{\pi}{2a} \leq k_{\mu} <  \frac{\pi}{2a}.
\end{equation}
The operators in (2.11) commute with $\gamma_{5}$, whereas those in (2.12) anti-commute with $\gamma_{5}$ and thus change the sign of  chiral charge,  reproducing the 15 species doublers with correct chiral charge assignment; $\sum_{n=0}^{15}(-1)^{n}\gamma_{5} =0$. 

In a smooth continuum limit, the operaor $\Cslash$ produces  $\Dslash$ for each species doubler with alternating chiral charge. The relation $Tr \gamma_{5} = 0$ or (2.3) for the operator $\Cslash$ is consistent  for any background gauge field because of the presence of these species doublers, which are degenerate with the physical species in the present case.

\subsection{Wilson operator $D_{W}$ and  $Tr\gamma_{5} = 0$}
The consistency of $Tr\gamma_{5} = 0$ is analyzed by means of topological properties which are  specified by  (2.8) and thus it is best described in the nearly continuum limit. 
To be more precise, one may define  the near continuum configurations by
the momentum $k_{\mu}$ carried by the fermion
\begin{equation}
- \frac{\pi}{2a}\epsilon \leq k_{\mu} \leq \frac{\pi}{2a}\epsilon
\end{equation}
for sufficiently small $a$ and $\epsilon$ combined with the operation ${\cal T}_{n}$ in (2.11) and (2.12). 
To identify each species doubler clearly in the near continuum configurations, we also keep $r/a$ and $m_{0}/a$ finite for $a\rightarrow$ small [14], and the gauge fields are assumed to be sufficiently smooth. For these configurations, we can approximate the operator $D_{W}$ by
\begin{equation}
D_{W}= i\Dslash + M_{n} + O(\epsilon^{2}) + O(agA_{\mu})
\end{equation}
for each species doubler, where the mass parameters $M_{n}$  stand for $M_{0}= - \frac{m_{0}}{a}$ and one of 
\begin{eqnarray}
&&\frac{2r}{a}-\frac{m_{0}}{a},\ \ (4,-1);\ \ \ 
\frac{4r}{a}-\frac{m_{0}}{a},\ \ (6,1)\nonumber\\
&&\frac{6r}{a}-\frac{m_{0}}{a},\ \ (4,-1);\ \ \ 
\frac{8r}{a}-\frac{m_{0}}{a},\ \ (1,1)
\end{eqnarray}
for $n=1\sim 15$. Here we denoted ( multiplicity, chiral charge ) in the bracket for species doublers. In (2.18) we used the relation valid for the configurations (2.17), for example, 
\begin{eqnarray}
D_{W}(k) &=& \sum_{\mu}\gamma^{\mu}\frac{\sin ak_{\mu}}{a} + \frac{r}{a}\sum_{\mu}(1 - \cos ak_{\mu}) - \frac{m_{0}}{a}\nonumber\\
&=& \gamma^{\mu}k_{\mu}( 1 + O(\epsilon^{2})) + \frac{r}{a} O(\epsilon^{2}) -
\frac{m_{0}}{a}
\end{eqnarray}
in the momentum representation with vanishing gauge field. 

In these near continuum configurations, the topological properties are specified by the operator  $\Dslash$  in $D_{W}$. We can thus evaluate $Tr\gamma_{5}$ by using the basis set defined by 
\begin{equation}
\Dslash\phi_{n} = \lambda_{n}\phi_{n}
\end{equation}
which formally {\em diagonalize} the effective  operator $D_{W}$ in (2.18) describing the low-energy excitations  of  each species doubler. We then obtain 
\begin{equation}
Tr \gamma_{5} = \sum_{n=0}^{15}(-1)^{n}\lim_{L\rightarrow large}\sum_{l=1}^{L}\phi_{l}^{\dagger}\gamma_{5}\phi_{l}=0
\end{equation}
where $\phi_{l}^{\dagger}\gamma_{5}\phi_{l}=0$ for $\lambda_{l}\neq 0$ because of $\gamma_{5}\Dslash + \Dslash\gamma_{5}=0$, and 
(2.22) states the cancellation of zero- mode contributions  $\sum_{\lambda_{l}=0}\phi_{l}^{\dagger}\gamma_{5}\phi_{l}$ among various species. We are assuming that our near continuum configurations (2.18) are  accurate in the treatment of these zero modes. An argument to support our identification of the near continuum configurations will be given in the next sub-section. 

$Tr\gamma_{5}=0$ is thus consistent even for a topologically non-trivial 
gauge background because of the presence of the would-be species doublers. This property is related to the well-known fact that one can safely ignore the Jacobian factor for global chiral transformation $\delta\psi = i\epsilon\gamma_{5}\psi$ and $\delta\bar{\psi}= \bar{\psi}i\epsilon\gamma_{5}$ for the theory defined by $S=\bar{\psi}D_{W}\psi$. 

\subsection{Overlap Dirac operator and  $Tr\gamma_{5} = 0$}

The operator $D$ introduced by Neuberger[2] , which satisfies the relation (1.1), has an explicit expression
\begin{equation}
aD= 1 - \gamma_{5}\frac{H}{\sqrt{H^{2}}} =1 + D_{W}\frac{1}{\sqrt{D_{W}^{\dagger}D_{W}}}
\end{equation}
where $D_{W}=-\gamma_{5} H$ is the Wilson operator.
For the near continuum configurations specified above in (2.17), one can approximate 
\begin{eqnarray}
D&=& \sum_{n=0}^{15}(1/a){[}1 + (i\Dslash + M_{n})\frac{1}{\sqrt{\Dslash^{2} + M_{n}^{2}}}]|n\rangle\langle n|,\nonumber\\
\gamma_{5}D&=& \sum_{n=0}^{15}(-1)^{n}\gamma_{5}(1/a){[}1 + (i\Dslash + M_{n})\frac{1}{\sqrt{\Dslash^{2} + M_{n}^{2}}}]|n\rangle\langle n|,\nonumber\\ 
\gamma_{5}&=& \sum_{n=0}^{15}(-1)^{n}\gamma_{5}|n\rangle\langle n|.
\end{eqnarray}
Here we explicitly write the projection  $|n\rangle\langle n|$  for each species doubler. The operators in (2.24) preserve the Ginsparg-Wilson relation (1.1).
 We can again use the basis set in (2.21), which formally diagonalize the basic
operator $D$ in (2.24), to define the trace operation. We thus obtain
\begin{equation} 
Tr \gamma_{5} = \sum_{n=0}^{15}(-1)^{n}\lim_{L\rightarrow large}\sum_{l=1}^{L}\phi_{l}^{\dagger}\gamma_{5}\phi_{l}=0
\end{equation}
by assuming that our effective operators (2.24) are accurate in describing the excitations near the zero modes $\Dslash\phi_{l}=0$ ,  which are relevant for 
topological considerations. Again the presence of the would-be species doublers makes the relation $Tr \gamma_{5}=0 $ consistent for any topologically non-trivial background gauge 
field. A justification of our effective description (2.24) will be given later.

The above expression of $D$ also shows that 
\begin{eqnarray}
D\phi_{l}& =& 0, \nonumber\\
D\phi_{l}&=& \frac{2}{a}\phi_{l}
\end{eqnarray}
for the physical species and the unphysical species doublers, respectively, if 
one uses the zero-modes $\Dslash\phi_{l}=0$.  Note that $M_{0}<0$ and 
the rest of $M_{n}>0$ in (2.19) and (2.24) [2]. We also note that $\phi_{l}$
can be a simultaneous eigenstate of $\gamma_{5}$ only for $\Dslash\phi_{l}=0$.  Namely, the $N_{\pm}$ states
with the eigenvalue $2/a$ in fact correspond to topological excitations associated with  species doublers; this means that the multiplicities of these $N_{\pm}$ are quite high  due to the 15 species 
doublers, although they  satisfy the sum rule  $n_{+} + N_{+} = n_{-} + N_{-}$.
This sum rule is a direct consequence of (2.25) and (2.26) by noting that $\phi_{l}^{\dagger}\gamma_{5}\phi_{l}= 0$ for $\lambda_{l}\neq 0$.

The  calculation of the index (1.2) may  proceed as 
\begin{eqnarray}
Tr \gamma_{5}(1 - \frac{a}{2}D) &=& - Tr\gamma_{5}\frac{a}{2}D\nonumber\\
&=& - \frac{1}{2} Tr \sum_{n=0}^{15}(-1)^{n}\gamma_{5}{[}1 + (i\Dslash + M_{n})\frac{1}{\sqrt{\Dslash^{2} + M_{n}^{2}}}]\nonumber\\
&=& - \frac{1}{2} Tr \sum_{n=0}^{15}(-1)^{n}\gamma_{5}(i\Dslash + M_{n})\frac{1}{\sqrt{\Dslash^{2} + M_{n}^{2}}}\nonumber\\  
&=& - \frac{1}{2}  \sum_{n=0}^{15}(-1)^{n}\sum_{l}\phi_{l}^{\dagger}\gamma_{5} M_{n}\frac{1}{\sqrt{\Dslash^{2} + M_{n}^{2}}}\phi_{l}\nonumber\\ 
&=& - \frac{1}{2}  \sum_{n=0}^{15}(-1)^{n}M_{n}\frac{1}{\sqrt{M_{n}^{2}}}
\sum_{\lambda_{l}=0}\phi_{l}^{\dagger}\gamma_{5}\phi_{l}\nonumber\\
&=& - \frac{1}{2}  \sum_{n=0}^{15}(-1)^{n}M_{n}\frac{1}{\sqrt{M_{n}^{2}}}(n_{+}
- n_{-})\nonumber\\
&=& - \frac{1}{2}(-1 + \sum_{n=1}^{15}(-1)^{n})(n_{+} - n_{-}) = n_{+} - n_{-}
\end{eqnarray}
where we used $\gamma_{5}\Dslash + \Dslash\gamma_{5}=0$ and the fact that $\phi_{l}^{\dagger}\gamma_{5}\phi_{l}= 0$ for $\lambda_{l}\neq 0$
 in $\Dslash\phi_{l}= \lambda_{l}\phi_{l}$. We also used the fact that $M_{0}<0$ and $M_{n}>0$ for $n=1\sim 15$ [2]. The index (2.27) is defined for $ \Dslash$ while the index (1.3) is defined for $\gamma_{5}D$, and both agree with  the Pontryagin index as is seen in (2.30) below. 

In the above calculation (2.27),  we used the relation $Tr \gamma_{5}= 0$ twice: In the second line, this relation requires the presence of the physical species as 
well as the species doublers. As a result, we have the contribution to the final index from both of the physical species and 15 species doublers, although the
species doublers with $\lambda_{l}= 2/a$ should saturate the index (and anomaly) in the expression[12]
\begin{equation}
- Tr (a/2)\gamma_{5}D =  n_{+} - n_{-}
\end{equation}
as is noted in (A.10) in Appendix. Our analysis of the global topological 
property on the basis of effective operators (2.24) is thus consistent. 

The above calculational scheme  of index (2.27)  in fact corresponds to the evaluation of the local index (i.e.,  anomaly)  performed in Ref.[8]. By using the plane wave basis , one has (in the limit $a\rightarrow 0$ with $r/a$ and $m_{0}/a$ kept fixed ) 
\begin{eqnarray}
&&tr \gamma_{5}(1 - \frac{a}{2}D)(x)\nonumber\\ 
&=& - \frac{1}{2} \sum_{n=0}^{15}(-1)^{n}tr \int^{\infty}_{-\infty} \frac{d^{4}k}{(2\pi)^{4}} e^{-ikx}\gamma_{5}(i\Dslash + M_{n})\frac{1}{\sqrt{\Dslash^{2} + M_{n}^{2}}}e^{ikx}\nonumber\\
&=& \frac{1}{2}tr \int^{\infty}_{-\infty} \frac{d^{4}k}{(2\pi)^{4}} e^{-ikx}\gamma_{5}\frac{1}{\sqrt{\Dslash^{2}/M_{0}^{2} + 1}}e^{ikx}\nonumber\\
&&-  \frac{1}{2}\sum_{n=1}^{15}(-1)^{n}tr \int^{\infty}_{-\infty} \frac{d^{4}k}{(2\pi)^{4}} e^{-ikx}\gamma_{5}\frac{1}{\sqrt{\Dslash^{2}/M_{n}^{2}+1}}e^{ikx}
\end{eqnarray}
which gives rise to the anomaly for all $|M_{n}| \rightarrow \infty$ in the continuum limit: 
\begin{eqnarray}
tr \gamma_{5}(1 - \frac{a}{2}D)(x)
&=& \lim_{M\rightarrow \infty}tr \int^{\infty}_{-\infty} \frac{d^{4}k}{(2\pi)^{4}} e^{-ikx}\gamma_{5}f(\frac{\Dslash^{2}}{M^{2}})e^{ikx}\nonumber\\
&=& \frac{g^{2}}{32\pi^{2}}tr \epsilon^{\mu\nu\alpha\beta}F_{\mu\nu}F_{\alpha\beta}. 
\end{eqnarray}
Here we defined $f(x)= 1/\sqrt{x+1}$ which satisfies 
\begin{eqnarray}
&& f(0)=1,\ \ f(\infty)=0,\nonumber\\
&& f^{\prime}(x)x|_{x=0} = f^{\prime}(x)x|_{x=\infty} = 0.
\end{eqnarray}
The right-hand side of (2.30) is known to be independent of the choice of
$f(x)$ which satisfies the mild condition (2.31)[11].

A direct evaluation of the anomaly without using (2.27) is of course possible. We briefly sketch the procedure here , since it  justifies our analysis based 
on the effective expressions in (2.24) ( and partly (2.18) also ). For an operator $O(x,y)$ defined on the lattice,
one may define 
\begin{equation}
O_{mn}\equiv \sum_{x,y}\phi_{m}^{\ast}(x)O(x,y)\phi_{n}(y),
\end{equation}
and the trace
\begin{eqnarray}
Tr O &=& \sum_{n}O_{nn}\nonumber\\
&=&\sum_{n}\sum_{x,y}\phi_{n}^{\ast}(x)O(x,y)\phi_{n}(y)\nonumber\\
&=&\sum_{x}(\sum_{n,y}\phi_{n}^{\ast}(x)O(x,y)\phi_{n}(y)).
\end{eqnarray}
The local version of the trace (or anomaly) is then defined by $tr O(x,x) \equiv\\ \sum_{n,y}\phi_{n}^{\ast}(x)O(x,y)\phi_{n}(y)$. For the operator of our interest, we have 
\begin{equation}
tr (-\frac{1}{2}\gamma_{5}D_{W}\frac{1}{\sqrt{D_{W}^{\dagger}D_{W}}})(x) =
-\frac{1}{2}\sum_{n=0}^{15} tr \int^{\frac{\pi}{2a}}_{-\frac{\pi}{2a}}\frac{d^{4}k}{(2\pi)^{4}}e^{-ikx}{\cal T}^{-1}_{n}\gamma_{5}D_{W}\frac{1}{\sqrt{D_{W}^{\dagger}D_{W}}}{\cal T}_{n}e^{ikx}
\end{equation}
where we used the plane wave basis defined in (2.16) combined with the operation
${\cal T}_{n}$. We also used a short hand notation $Oe^{ikx}= \sum_{y}O(x,y)e^{iky}$.

We first take the  $a\rightarrow 0$ limit of this expression with all $M_{n}, 
n=0\sim 15$, kept fixed and then take the limit $|M_{n}| \rightarrow \infty$
later. For fixed $M_{n}$ ( to be precise, for fixed $m_{0}/a$ and $r/a$ ),
one can confirm that the above integral (2.34) for the domain $\frac{\pi}{2a}\epsilon \leq |k_{\mu}| \leq \frac{\pi}{2a}$ vanishes ( at least ) linearly in $a$ for $a \rightarrow 0$, if one takes into account the trace with $\gamma_{5}$. See also Refs.[7][9]. In the remaining integral 
\begin{equation}
-\frac{1}{2}\sum_{n=0}^{15} (-1)^{n}tr \int^{\frac{\pi}{2a}\epsilon}_{-\frac{\pi}{2a}\epsilon}\frac{d^{4}k}{(2\pi)^{4}}e^{-ikx}\gamma_{5}{\cal T}^{-1}_{n}D_{W}\frac{1}{\sqrt{D_{W}^{\dagger}D_{W}}}{\cal T}_{n}e^{ikx}
\end{equation}
one may take the limit $a\rightarrow 0$ ( and $\frac{\pi}{2a}\epsilon \rightarrow \infty$ ) with letting $\epsilon$ arbitrarily small. By taking (2.20) into 
account, one thus recovers the  expression (2.29). One can arrive at the same 
conclusion by using an auxiliary 
regulator $h(\Cslash^{2}/m^{2})$ in the integrand in (2.34) to make the intermediate steps better defined[8]. The domain in (2.17) with arbitrarily small but finite $\epsilon$  thus correctly describes the topological aspects of the continuum limit in the present prescription. 

Here we went through the details of the anomaly calculation to show that the interpretation of the 
$N_{\pm}$ states in (A.8) as topological excitations related to 
species doublers, as is shown in (2.26), is also consistent with the  local 
anomaly calculation. As for a general analysis of chiral anomaly
in the overlap operator, see Ref.[15].

At this stage it is instructive to consider an operator defined by 
\begin{equation}
D\equiv \frac{1}{a}[ 1 + (i\Dslash + M_{0})\frac{1}{\sqrt{\Dslash^{2} + M_{0}^{2}}}]
\end{equation}
instead of $D$ in (2.24). This $D$ is regarded as an $M_{n}\rightarrow\infty, \
n\neq 0$, limit of the effective operator $D$ (2.24) in the Lagrangian level, and it satisfies ( a continuum version of ) the Ginsparg-Wilson relation (1.1) without any species doubler. The relation  
$Tr \gamma_{5}=0$ is thus expected to be inconsistent. In fact we have an index related to the chiral Jacobian [5] 
\begin{eqnarray}
Tr \gamma_{5}( 1 - \frac{a}{2}D) &=& \lim_{L\rightarrow large}\sum_{l=1}^{L}\phi_{l}^{\dagger}\gamma_{5}( 1 - \frac{a}{2}D)\phi_{l}\nonumber\\
&=& \sum_{\lambda_{l}=0}\phi_{l}^{\dagger}\gamma_{5}\phi_{l} = n_{+} - n_{-}
\end{eqnarray}
by noting $\phi_{l}^{\dagger}\gamma_{5}\phi_{l}=0$ for $\lambda_{l}\neq 0$ in 
$\Dslash\phi_{l}=\lambda_{l}\phi_{l}$. On the other hand, if one incorrectly
uses $Tr \gamma_{5}=0$ one obtains 
\begin{eqnarray}
Tr \gamma_{5}( 1 - \frac{a}{2}D) &=& -\frac{1}{2}Tr ( \gamma_{5}(i\Dslash + M_{0})\frac{1}{\sqrt{\Dslash^{2} + M_{0}^{2}}})\nonumber\\
&=& -\frac{1}{2}\lim_{L\rightarrow large}\sum_{l=1}^{L}\phi^{\dagger}_{l} \gamma_{5}M_{0}\frac{1}{\sqrt{\Dslash^{2} + M_{0}^{2}}}\phi_{l}\nonumber\\
&  =& \frac{1}{2}(n_{+} - n_{-})
\end{eqnarray}
by noting $\gamma_{5}\Dslash + \Dslash\gamma_{5} =0$,  $\phi_{l}^{\dagger}\gamma_{5}\phi_{l}=0$ for $\lambda_{l}\neq 0$, and $M_{0}<0$. One thus looses half of the index or anomaly. In this example, the evaluation of $Tr \gamma_{5}$ is somewhat subtle, but $Tr \gamma_{5}=0$ is definitely inconsistent since the calculation in the last line in (2.38) is well-defined. In fact the relations 
\begin{equation}
Tr \gamma_{5} = n_{+} - n_{-}\ \ and \ \ Tr (-\frac{a}{2}\gamma_{5}D) = 0
\end{equation}
are consistent for the present operator $D$, since the species doublers at
$\gamma_{5}D\phi_{l} = \pm (2/a)\phi_{l}$ are missing. A more rigorously  regularized Jacobian for the present example is given by the formula (3.3) to be discussed later. 
 
\subsection{General lattice Dirac operator and $Tr \gamma_{5}=0$}
We expect that our analysis of $Tr \gamma_{5}=0$, namely its consistency is ensured only by the presence of the would-be species doublers in the Hilbert space,  works for a general lattice Dirac operator, since any lattice operator contains $\Cslash$ as an essential part. For the smooth near continuum configurations, the lowest dimensional operator $\Cslash$ is expected to specify the topological properties. From this viewpoint, the overlap Dirac operator $D$ describes the topological properties such as the index theorem and $Tr \gamma_{5}=0$ in a neater way than the Wilson operator $D_{W}$, mainly because the operator $D$ projects all the species 
doublers to the vicinity of $2/a$: The behavior for small values of $\Cslash$ (i.e.,for  $|\Cslash| \ll 1/a$ )  is described in a more clear-cut way by $D$,
and one can recognize clearly the topological $N_{\pm}$ states related to species doublers. 

We here note that the Pauli-Villars regularization in continuum theory can be 
analyzed in a similar way. The Pauli-Villars regulator is defined in the path 
integral by introducing a bosonic spinor $\phi$ into the action
\begin{equation}
S = \int d^{4}x [\bar{\psi}(i\Dslash - m )\psi + \bar{\phi}(i\Dslash - M )
\phi ].
\end{equation}
The Jacobian for the global chiral transformation then gives rise to the graded
trace[11]
\begin{equation}
Tr \gamma_{5} = Tr_{\psi}\gamma_{5} - Tr_{\phi}\gamma_{5} = 0.
\end{equation}
The relation $Tr \gamma_{5} = 0$ is thus consistent with any topologically non-trivial background  gauge field because of the presence of the unphysical regulator $\phi$. This $\phi$ is analogous to the species doublers in lattice regularization. 

\section{Implications of the present analysis}
We have shown that the consistency of $Tr \gamma_{5} = 0$ for topologically non-trivial background gauge fields requires the presence
 of some unphysical states in the Hilbert space. Coming back to the original lattice theory defined by 
\begin{equation}
S = \bar{\psi}D\psi
\end{equation}
with $D$ satisfying the relation (1.1), one obtains twice of  (1.3) as a Jacobian factor for the global chiral transformation [5] $\delta\psi = i\epsilon\gamma_{5}(1-\frac{a}{2}D)\psi$ and $\delta\bar{\psi} = \bar{\psi}i\epsilon (1-\frac{a}{2}D)\gamma_{5}$, which leaves the action (3.1) invariant. One can rewrite (1.3) as 
\begin{equation}
Tr \gamma_{5}( 1 - \frac{a}{2}D) = \tilde{T}r \gamma_{5}( 1 - \frac{a}{2}D) =
\tilde{T}r \gamma_{5} = n_{+} - n_{-}
\end{equation}
where the modified trace $\tilde{T}r$ is defined  by truncating the unphysical 
$N_{\pm}$ states with $\lambda_{n} =\pm  2/a$ . Without the $N_{\pm}$ states, $\tilde{T}r \gamma_{5}\frac{a}{2}D = 0$ since the eigenvalues $\lambda_{n}$ of $\gamma_{5}D$ with $\lambda_{n} \neq 0, \pm 2/a$ appear always pairwise at $\pm |\lambda_{n}|$. See Appendix.

If one takes a smooth continuum limit of $\tilde{T}r \gamma_{5}= n_{+} - n_{-}$ in (3.2) , one recovers the result of the continuum path integral (1.6). If one considers that $\tilde{T}r \gamma_{5}$ is too abstract, one may  define it more concretely by
\begin{eqnarray}
Tr \gamma_{5}( 1 - \frac{a}{2}D) f(\frac{(\gamma_{5}D)^{2}}{M^{2}})&=& \tilde{T}r \gamma_{5}( 1 - \frac{a}{2}D)f(\frac{(\gamma_{5}D)^{2}}{M^{2}})\nonumber\\
&=& \tilde{T}r \gamma_{5}f(\frac{(\gamma_{5}D)^{2}}{M^{2}}) = n_{+} - n_{-}
\end{eqnarray}
for {\em any } $f(x)$ which satisfies the mild condition in (2.31). See also (1.3) and (1.5). This relation 
suggests that we can extract the local index ( or anomaly) by  
\begin{equation}
tr \gamma_{5}( 1 - \frac{a}{2}D) f(\frac{(\gamma_{5}D)^{2}}{M^{2}})(x)  
\end{equation}
which  is shown to be independent of the choice of $f(x)$ in the limit $a \rightarrow 0$ and leads to (2.30) ( for $f(x)$ which goes to zero rapidly for $x\rightarrow \infty$ ) by using only the general properties of $D$ [8]. If one constrains the momentum domain to (2.16) from the beginning , one may  use the last expression in (3.3) to evaluate the anomaly for a more general class of $f(x)$.  We thus naturally recover the result of the continuum path integral[11]. 

As for a more practical implication of our analysis of $Tr \gamma_{5} = 0$ in  lattice theory, one may say that any  result which depends  critically on  the states $N_{\pm}$  is  {\em unphysical}. It is thus necessary to define the scalar density ( or mass term ) and pseudo-scalar density in the theory (3.1) by [16][6]
\begin{eqnarray}
S(x) &=& \bar{\psi}_{L}\psi_{R} + \bar{\psi}_{R}\psi_{L}= \bar{\psi}( 1 - 
\frac{a}{2}D)\psi\nonumber\\
P(x) &=& \bar{\psi}_{L}\psi_{R} - \bar{\psi}_{R}\psi_{L}= \bar{\psi}\gamma_{5}( 1 - \frac{a}{2}D)\psi
\end{eqnarray}
Here we defined two independent projection operators 
\begin{eqnarray}
P_{\pm} &=& \frac{1}{2}( 1 \pm \gamma_{5})\nonumber\\
\hat{P}_{\pm} &=& \frac{1}{2}( 1 \pm \hat{\gamma}_{5})   
\end{eqnarray}
with $\hat{\gamma}_{5} = \gamma_{5}(1 - aD)$ which satisfies $\hat{\gamma}_{5}^{2} = 1$ [6]. The left- and right- components are then defined by 
\begin{equation}
\bar{\psi}_{L,R}= \bar{\psi}P_{\pm}, \ \ \psi_{R,L}= \hat{P}_{\pm}\psi
\end{equation}
which is based on the decomposition
\begin{equation}
D = P_{+}D\hat{P}_{-} +  P_{-} D\hat{P}_{+}.    
\end{equation}
The physical operators $S(x)$ and $P(x)$ in (3.5) do not contain the contribution from the unphysical
states $N_{\pm}$ in (A.8). In the spirit of  this construction, the definition of the index  by (3.3) which is independent of unphysical states $N_{\pm}$ is natural.
 In particular, all the unphysical species doublers ( not only the topological
ones at $2/a$ ) decouple from the anomaly defined by (3.4) in the limit 
$a \rightarrow 0$ with fixed $M$.  

The customary calculation of the index ( and also anomaly ) by the 
relation[4]-[7][9]
\begin{equation}
Tr \gamma_{5}( 1 - \frac{a}{2}D) = Tr( - \frac{a}{2}\gamma_{5}D) = n_{+} - n_{-} 
\end{equation}
by itself is of course consistent, since one simply includes the unphysical states $N_{\pm}$ in evaluating  $Tr \gamma_{5} = 0$, and consequently one obtains the index 
$Tr ( - \frac{a}{2}\gamma_{5}D)$ from the unphysical states $N_{\pm}$ only.
We after all know that the left-hand side of (3.9) is independent of $N_{\pm}$. 

Rather, the major message of our analysis is that the continuum limit of 
$Tr \gamma_{5} = 0$ in (1.8) ( unlike the relation $\tilde{T}r \gamma_{5} = n_{+} - n_{-}$ ) {\em cannot } be defined in a consistent way when the (would-be) species doublers disappear from the Hilbert space. It is clear from the expression of $Tr \gamma_{5} = 0$ in (1.8) that the $a\rightarrow 0$ limit of $Tr \gamma_{5} = 0$ is not defined consistently. One may then ask how the calculation of local anomaly on the basis of (3.9) could be consistent in the limit $a\rightarrow 0$ if $Tr \gamma_{5} = 0$ is inconsistent?  A key to resolve this apparent paradox is the failure of the decoupling of heavy fermions in the evaluation of anomaly. The massive unphysical species doublers do not decouple from the anomaly , as is seen in (2.29), for example. If one insists on $Tr \gamma_{5} = 0$ in the continuum limit, one is also insisting on the failure of the decoupling of these infinitely massive particles from $Tr \gamma_{5} = 0$. The contributions of these heavy fermions to the anomaly!
 and to $Tr \gamma_{5} = 0$ precisely cancel, just as in the case of the evaluation of global index in (3.9). Namely, the local anomaly itself is {\em independent} of these massive species doublers in the  continuum limit, as is clear in (3.4). In this sense, (3.4) is the only logically consistent definition of local anomaly. It is an advantage of the finite lattice formulation that we can now clearly illustrate this subtle cancellation of the contributions of those ultra-heavy regulators to  $Tr \gamma_{5} = 0$ and 
anomaly on the basis of (1.8). ( In the case of the Wilson fermion operator 
$D_{W}$, an analogous cancellation takes place in $Tr \gamma_{5} + pseudo-scalar\ \  mass\ \  term$ induced by the chiral variation of the action.) 

When one defines a chiral theory by recalling (3.8) [6][17] 
\begin{equation}
S =  \bar{\psi} P_{+}D\hat{P}_{-}\psi =  \bar{\psi}_{L} D \psi_{L}
\end{equation}
one obtains the {\em covariant} gauge anomaly (or Jacobian)
\begin{equation}
tr T^{a}\gamma_{5}( 1 - \frac{a}{2}D) = \sum_{n}\phi_{n}(x)^{\dagger}T^{a}\gamma_{5}( 1 - \frac{a}{2}D)\phi_{n}(x)
\end{equation}
for the gauge transformation $\delta\psi_{L}(x) = i\alpha^{a}(x)T^{a}\hat{P}_{-}\psi_{L}$  and  $\delta\bar{\psi}_{L}(x) = \bar{\psi}_{L}P_{+}(-i)\alpha^{a}(x)T^{a}$. 

An  analogue of the $U(1)$ anomaly  (3.4) is then defined for the gauge anomaly (3.11) by ( by using $\phi_{n}$ in (1.4))
\begin{equation}
\sum_{n}\phi_{n}(x)^{\dagger}T^{a}\gamma_{5}( 1 - \frac{a}{2}D)f(\frac{(\gamma_{5}D)^{2}}{M^{2}})\phi_{n}(x) = \sum_{n}f(\frac{\lambda_{n}^{2}}{M^{2}})\phi_{n}(x)^{\dagger}T^{a}(\gamma_{5} - \frac{a}{2}\lambda_{n})\phi_{n}(x)
\end{equation}
which reduces to the lattice expression for $M \rightarrow \infty$  with $f(0)=1$. In practice, one first takes the continuum limit $ a \rightarrow 0$ with $M$ fixed and one obtains (see , for example, [8])
\begin{equation}
tr T^{a}\gamma_{5}f(\frac{\Dslash^{2}}{M^{2}})
\end{equation}
which is again known to be independent of the specific choice of $f(x)$  in the limit
$M \rightarrow \infty$[11]. 
For the overlap Dirac operator, one can show that the anomaly calculation in (3.11) by using $tr T^{a}\gamma_{5}=0$ corresponds effectively to a specific choice of $f(x) = 1/\sqrt{ 1 + x}$ in (3.13), just as in the case of  $U(1)$ anomaly in (2.30). 

The definition of the regularized Jacobian (3.12) may be regarded to  correspond to the truncation of  the states $N_{\pm}$  from the chiral action
\begin{equation}
S = \sum_{n\in N_{+}} (\frac{2}{a})\bar{C}_{n}C_{n} + \sum_{0\leq\lambda_{n}< 2/a} \lambda_{n}\bar{C}_{n}C_{n}
\end{equation}
to 
\begin{equation}
\tilde{S} = \sum_{0\leq\lambda_{n}< 2/a}\lambda_{n}\bar{C}_{n}C_{n}
\end{equation} 
and then taking the continuum limit $ a \rightarrow 0$, which is logically
more natural  as the $N_{\pm}$ states are eliminated from the Hilbert space 
{\em before} taking the continuum limit. 

Incidentally, the action (3.14) is obtained from (3.10) by expanding
\begin{eqnarray}
\bar{\psi}P_{+} &=& \sum _{n} \bar{C}_{n}\bar{v}_{n},\nonumber\\
\hat{P}_{-}\psi &=& \sum _{n} C_{n} v_{n}  
\end{eqnarray}
with the choice of the basis set 
\begin{eqnarray}
\{ v_{j}\} &=& \{ \phi_{n}| \gamma_{5}D\phi_{n}= 0, \gamma_{5}\phi_{n}= - \phi_{n}\}\nonumber\\ &\oplus&   \{ \phi_{n}| \gamma_{5}D\phi_{n}= 2/a\phi_{n}, \gamma_{5}\phi_{n}= + \phi_{n}\}\nonumber\\
 &\oplus&  \{ \hat{P}_{-}\phi_{n}/\sqrt{(1+a\lambda_{n}/2)/2}\ | \gamma_{5}D\phi_{n}= \lambda_{n}\phi_{n}, 2/a > \lambda_{n} > 0 \}
\end{eqnarray}
\begin{eqnarray}
\{ \bar{v}_{k}^{\dagger}\} &=& \{ \phi_{n}| \gamma_{5}D\phi_{n}= 0, \gamma_{5}\phi_{n}= + \phi_{n}\}\nonumber\\ & \oplus&   \{ \phi_{n}| \gamma_{5}D\phi_{n}= 2/a\phi_{n}, \gamma_{5}\phi_{n}= + \phi_{n}\}\nonumber\\
 &\oplus&  \{ P_{+}\phi_{n}/\sqrt{(1+a\lambda_{n}/2)/2}\ | \gamma_{5}D\phi_{n}= \lambda_{n}\phi_{n}, 2/a > \lambda_{n} > 0 \}
\end{eqnarray}
in terms of the eigenstates of $\gamma_{5}D\phi_{n} = \lambda_{n}\phi_{n}$ summarized in Appendix.

Consequently, the path integral for a fixed background gauge field is defined 
by
\begin{eqnarray}
Z &=& J \int \prod_{n\in N_{+}}d\bar{C}_{n}dC_{n}\prod_{0\leq\lambda_{n}< 2/a}
d\bar{C}_{n}\prod_{0\leq\lambda_{m}< 2/a}dC_{m}\exp S, \nonumber\\
\tilde{Z} &=& \tilde{J}\int \prod_{0\leq\lambda_{n}< 2/a}
d\bar{C}_{n}\prod_{0\leq\lambda_{m}< 2/a}dC_{m}\exp \tilde{S}
\end{eqnarray} 
with Jacobian factors $J$ and $\tilde{J}$ which depend on the basis set.
A ( naive ) continuum limit of the truncated expression $\tilde{Z}$ naturally gives rise to the covariant path integral formulation of  chiral gauge theory[11]. In particular, the fermion number anomaly which is given by (3.3) gives rise to the fermion number violation in chiral gauge thoery.  As is well-known, this formulation of the continuum limit is consistent if the anomaly cancellation condition $tr T^{a}\{ T^{b},T^{c}\} = 0$ is 
satisfied, when combined with the argument of the robustness of lattice gauge symmetry [18]. See also [19].   

An interesting analysis of the definition of chiral theory at a finite $a$ has been given  by L\"{u}scher recently [20]. The fermion number violation arises from the non-trivial index of the rectangular ( {\em not} square) matrix in (3.10) (Cf. (2.3))
\begin{equation}
dim\ \ ker\ \ \hat{P}_{-}\gamma_{5}DP_{+} - dim\ \ ker\ \ P_{+}\gamma_{5}D\hat{P}_{-} = n_{+} - n_{-}
\end{equation}
as is seen in the explicit construction of the basis vectors in (3.17) and (3.18): For a general $n\times m$ matrix $M$, one can prove an index theorem 
\begin{equation}
dim\ ker\ M - dim\ ker\ M^{\dagger} = m - n
\end{equation}
which is a generalization  of the case of a square matrix with $m=n$. For the operator $ \hat{P}_{-}\gamma_{5}DP_{+}$ in (3.20), the dimensions of the column and row vectors are respectively given by using the projection operators as 
$Tr \hat{P}_{-}$ and $Tr P_{+}$, and thus $ m - n = Tr P_{+} - Tr \hat{P}_{-}
= Tr \gamma_{5}(1-\frac{a}{2}D) = n_{+} - n_{-}$ [20]. 
Incidentally, an analogous analysis provides an alternative proof of the equivalence of $Tr\gamma_{5}=0$ with the index relation (2.3).
 
\section{Discussion and conclusion}
Motivated by the recent interesting developments in lattice gauge theory,  we analyzed the physical implications of the condition $Tr \gamma_{5} = 0$ in detail.  We have shown that  $Tr \gamma_{5} = 0$, whose validity is often taken for granted,  is consistent only when one includes some unphysical states in the Hilbert space.  The continuum $a\rightarrow 0$ limit of $Tr \gamma_{5} = 0$ is not defined consistently as is seen in (1.8). We have explained  that the failure of the decoupling of heavy fermions in the anomaly calculation  is a key to understand the consistency of the customary lattice calculation  of anomaly where $Tr \gamma_{5} = 0$ is used. Our analysis is perfectly consistent with the relation  (1.6) in the continuum path integral and even provides  positive support for the formula (3.3) and the related definition of anomaly (3.4) in lattice theory.

We here want to comment on an analysis of the photon phase operator[21] where a closely related phenomenon associated with the notion of index takes place [22]. The Maxwell field is expanded into an infinite set of harmonic oscillators, and thus the analysis of the photon phase operator is performed for a simple harmonic oscillator
\begin{equation}
H = \frac{1}{2}( p^{2} + \omega^{2}q^{2}) = \hbar \omega ( a^{\dagger}a + \frac{1}{2})
\end{equation} 
The quantum requirement  of the absence of the negative normed states leads to 
$a|0\rangle = 0$, and thus the index relation  
\begin{equation}
dim \ \ ker \ \ a - dim \ \ ker \ \ a^{\dagger} = 1
\end{equation}
since no states are annihilated by $ a^{\dagger}$. On the other hand, the existence of the observable {\em hermitian} phase operator $\varphi$ requires a decomposition [21]
\begin{equation}
a = U(\varphi)\sqrt{N}, \ \ \ a^{\dagger} = \sqrt{N}U(\varphi)^{-1}
\end{equation}
with a {\em unitary} $U(\varphi) = e^{i\varphi}$ and $N = a^{\dagger}a$. These expressions  suggest 
\begin{equation}
dim \ \ ker \ \ a - dim \ \ ker \ \ a^{\dagger} = 0
\end{equation}
in contradiction to the relation (4.2), since the unitary factor $U(\varphi)$ does not influence the analysis of index. The index (4.2) thus provides a no go theorem against  the hermitian photon phase operator and the resulting familiar phase-number uncertainty relation [22]. 

To circumvent the topological stricture (4.2), one may truncate the operator $a$ to an $(s+1)\times (s+1)$ dimensional square matrix 
\begin{eqnarray}
a_{s}&=&\left(\begin{array}{cccccc}
          0&1&0    &0    &..&0\\
          0&0    &\sqrt{2}&0    &..&0\\
          0&0    &0    &\sqrt{3}&..&0\\
          .&.&.&.    &.        &.\\
          0&0    &0    &0&..&\sqrt{s}    \\
          0&0    &0    &0&..&0
          \end{array}\right)\nonumber\\
&=& |0\rangle\langle 1| + |1\rangle\langle 2|\sqrt{2} ....  + |s-1\rangle\langle s|\sqrt{s}
\end{eqnarray}
and $a_{s}^{\dagger} = (a_{s})^{\dagger}$. One then obtains a vanishing index for a finite dimensional square matrix [22]
\begin{equation}
dim \ \ ker \ \ a_{s} - dim \ \ ker \ \ a_{s}^{\dagger} = 0
\end{equation}
and one can in fact introduce a hermitian phase operator $\phi$ [23]  which satisfies the relation $a_{s}= e^{i\phi}\sqrt{a_{s}^{\dagger}a_{s}}$.

The parameter $s$ or the state $|s\rangle$ stands for the cut-off parameter  analogous to the $N_{\pm}$ states related to $Tr \gamma_{5} = 0$ in lattice theory. A careful analysis of the uncertainty relation shows that the hermitian operator $\phi$ , when used to analyze the data which is already in the quantum limit, leads to a substantial deviation from the minimum uncertainty relation at the characteristically quantum domain with {\em small} average photon numbers. This artificial deviation from the minimum uncertainty is caused by the presence of the unphysical cut-off introduced by $|s\rangle$,  which fails to decouple from the low energy quantities for arbitrarily large but finite $s$ [22]. Also,
a large $s$ limit of (4.6) is not defined consistently, which is analogous to
the ill-defined continuum limit of $Tr \gamma_{5}= 0$ in (1.8).

It is expected that an analogous  unphysical result will appear in lattice gauge theory if one analyzes the low energy quantity which critically depends on the unphysical states $N_{\pm}$. In fact , it is known that one {\em has to } eliminate the contribution of the $N_{\pm}$ states to the physical observables such as $S(x)$ and $P(x)$ in (3.5) [16][6].

\appendix

\section{Finite dimensional representations of the Ginsparg - Wilson algebra}

In this Appendix  we recapitulate  the finite dimensional representations of the basic algebraic relation (1.1). A construction of the operator $\gamma_{5}D$, which satisfies the Ginsparg-Wilson relation on a finite lattice, by using  a corresponding  operator $\gamma_{5}D$ on an 
infinite lattice has been discussed in Ref.[20]. We first  define an operator
\begin{equation}
\Gamma_{5}\equiv \gamma_{5}(1-\frac{1}{2}aD)
\end{equation}
which is hermitian and satisfies the basic relation
\begin{equation}
\Gamma_{5}\gamma_{5}D +  \gamma_{5}D\Gamma_{5}=0.
\end{equation}
This relation suggests that if
\begin{equation}
\gamma_{5}D\phi_{n} = \lambda_{n}\phi_{n}, \ \ \ (\phi_{n},\phi_{n}) =1 
\end{equation}
then
\begin{equation}
\gamma_{5}D(\Gamma_{5}\phi_{n}) = -\lambda_{n}(\Gamma_{5}\phi_{n}).
\end{equation}
Namely, the eigenvalues $\lambda_{n}$ and $-\lambda_{n}$ are always paired 
if $\lambda_{n}\neq 0$ and $(\Gamma_{5}\phi_{n},\Gamma_{5}\phi_{n})\neq 0$.

We  evaluate the norm of $\Gamma_{5}\phi_{n}$
\begin{eqnarray}
(\Gamma_{5}\phi_{n},\Gamma_{5}\phi_{n})
&=& (\phi_{n},(\gamma_{5} - \frac{a}{2}\gamma_{5}D)(\gamma_{5} - \frac{a}{2}\gamma_{5}D)\phi_{n})\nonumber\\ 
&=&(\phi_{n},(1-  \frac{a}{2}\gamma_{5}(\gamma_{5}D+ D\gamma_{5}) + \frac{a^{2}}{4}(\gamma_{5}D)^{2})\phi_{n})\nonumber\\
&=&(\phi_{n},(1-  \frac{a^{2}}{4}(\gamma_{5}D)^{2})\phi_{n})\nonumber\\
&=&(1-  \frac{a}{2}\lambda_{n})(1+ \frac{a}{2}\lambda_{n}).
\end{eqnarray}
Namely $\phi_{n}$ is a ``highest'' state 
\begin{equation}
\Gamma_{5}\phi_{n}=(\gamma_{5} - \frac{a}{2}\gamma_{5}D)\phi_{n}=0
\end{equation}
if $(1-  \frac{a}{2}\lambda_{n})(1+ \frac{a}{2}\lambda_{n})=0 $
for the Euclidean $SO(4)$- invariant positive definite inner product 
$(\phi_{n}, \phi_{n})$.
We thus conclude that the states $\phi_{n}$ with $\lambda_{n}= \pm \frac{2}{a}$
 are {\em not} paired by the operation $\Gamma_{5}\phi_{n}$ and are the 
simultaneous eigenstates of $\gamma_{5}$, $\gamma_{5}\phi_{n}= \pm \phi_{n} $
respectively. One can also show that these eigenvalues $\lambda_{n}$ are the 
maximum or minimum of the possible eigenvalues of $\gamma_{5}D$. This is based
on the relation (1.5),  $ |\frac{a\lambda_{n}}{2}| = |\phi^{\dagger}_{n}\gamma_{5}\phi_{n}|\leq ||\phi_{n}|| ||\gamma_{5}\phi_{n}|| = 1 $.

On the other hand, the relation $Tr \gamma_{5}=0$, which is expected to be valid on a finite lattice  leads to ( by using (1.5))
\begin{eqnarray}
Tr\gamma_{5} &=& \sum_{n} \phi^{\dagger}_{n}\gamma_{5}\phi_{n}\nonumber\\
&=& \sum_{ \lambda_{n}=0}\phi^{\dagger}_{n}\gamma_{5}\phi_{n} +
\sum_{ \lambda_{n}\neq 0}\phi^{\dagger}_{n}\gamma_{5}\phi_{n}\nonumber\\
&=& \sum_{ \lambda_{n}=0}\phi^{\dagger}_{n}\gamma_{5}\phi_{n}
+ \sum_{ \lambda_{n}\neq 0}\frac{a}{2}\lambda_{n}\nonumber\\
&=& n_{+} - n_{-} +  \sum_{ \lambda_{n}\neq 0}\frac{a}{2}\lambda_{n}=0.
\end{eqnarray}
In the last line  of this relation, all the states except for the 
states with $\lambda_{n}= \pm 2/a$  cancel pairwise for $\lambda_{n}\neq
0$. We thus obtain a chirality  sum rule $ n_{+} - n_{-} + N_{+} -  N_{-} =0 
$ [12] or,  
\begin{equation}
n_{+}+ N_{+} =  n_{-} + N_{-}  
\end{equation} 
where $N_{\pm}$ stand for the number of isolated (un-paired) states with $\lambda_{n}=
\pm 2/a$ and $\gamma_{5}\phi_{n} = \pm \phi_{n}$, respectively. These relations show that the chirality asymmetry at vanishing eigenvalues is balanced by the chirality asymmetry at the largest  eigenvalues with $|\lambda_{n}|= 2/a$. 

We note that all other states with $0 < |\lambda_{n}| < 2/a$, which appear pairwise with $\lambda_{n}= \pm |\lambda_{n}|$ ( note that $\Gamma_{5}(\Gamma_{5}\phi_{n}) = (1 - (a\lambda_{n}/2)^{2})\phi_{n} \propto \phi_{n}$ for 
$|a\lambda_{n}/2|\neq 1$ )
, satisfy the relations
\begin{eqnarray}
\phi^{\dagger}_{n}\Gamma_{5}\phi_{n}&=&0,\nonumber\\
\phi^{\dagger}_{n}\gamma_{5}\phi_{n}&=& \frac{a\lambda_{n}}{2},\nonumber\\
\phi^{\dagger}_{m}\gamma_{5}\phi_{n}&=& 0 \ \ for \ \ \lambda_{m}\neq \lambda_{n}, \ \ \lambda_{m}\lambda_{n}>0.
\end{eqnarray}
These states $\phi_{n}$ cannot be the eigenstates of $\gamma_{5}$ as 
$|a\lambda_{n}/2| < 1$. 
The states $N_{\pm}$ saturate the index theorem commonly written in the form [4]-[6]
\begin{equation}
 Tr( \frac{-1}{2}a\gamma_{5}D) = n_{+} - n_{-}
\end{equation}
namely, only the states $N_{\pm}$ contribute to the left-hand side.

Those properties we analyzed so far in this Appendix  hold both for non-Abelian and Abelian gauge theories. We did not specify precise boundary conditions, since our analysis is valid once non-trivial zero modes appear for a given boundary condition. For an Abelian theory, one needs to introduce the gauge field configuration  with suitable boundary conditions, which carries a  non-vanishing magnetic flux , to generate a non-trivial index $n_{+} - n_{-}$ [20]. Our analysis of the index in this Appendix  is  formal, since it is well known that 
the Ginsparg-Wilson relation (1.1) by itself does not uniquely specify the index or the coefficient of chiral anomaly for a given gauge field configuration [24].  

To summarize the analyses of the present Appendix, all the normalizable eigenstates $\phi_{n}$ of $\gamma_{5}D$ on a finite lattice are categorized into the following 3 classes:\\
(i)\ $n_{\pm}$ states,\\
\begin{equation}
\gamma_{5}D\phi_{n}=0, \ \ \gamma_{5}\phi_{n} = \pm \phi_{n},
\end{equation}
(ii)\ $N_{\pm}$ states, \\
\begin{equation}
\gamma_{5}D\phi_{n}= \pm \frac{2}{a}\phi_{n}, \ \ \gamma_{5}\phi_{n} = \pm \phi_{n},\ \ respectively,
\end{equation}
(iii) Remaining states with $0 < |\lambda_{n}| < 2/a$,
\begin{equation}
\gamma_{5}D\phi_{n}= \lambda_{n}\phi_{n}, \ \ \ \gamma_{5}D(\Gamma_{5}\phi_{n})
= - \lambda_{n}(\Gamma_{5}\phi_{n}), 
\end{equation}
and the sum rule $n_{+}+ N_{+} =  n_{-} + N_{-}$ holds. 

All the $n_{\pm}$ and 
$N_{\pm}$ states are the eigenstates of $D$,  $D\phi_{n}=0$ and $D\phi_{n}= (2/
a) \phi_{n}$, respectively. If one denotes the number of states  in (iii) 
by $2N_{0}$, the total number of states $N$ is given by $N = 2(n_{+} + N_{+} +
N_{0})$, which is expected to be a constant independent of background gauge 
field configurations.

\end{document}